\documentclass[conference]{IEEEtran}

\usepackage[pdftex]{graphicx}
\usepackage{array}
\usepackage{amsfonts}
\usepackage{amsmath}
\usepackage{tikz} 
\usepackage{caption,subcaption}

\def\P{\mathbb{P}}
\def\E{\mathbb{E}}

\begin{document}
%
\title{Percolation for D2D Networks on Street Systems}

\author{\IEEEauthorblockN{
Elie Cali\IEEEauthorrefmark{2},
Nila Novita Gafur\IEEEauthorrefmark{1},
Christian Hirsch\IEEEauthorrefmark{3}, 
Benedikt Jahnel\IEEEauthorrefmark{4} 
Taoufik En-Najjary\IEEEauthorrefmark{2} and 
Robert I.A.~Patterson\IEEEauthorrefmark{4}}
\\
\IEEEauthorblockA{\IEEEauthorrefmark{2}Orange Labs Network, 44 Avenue de la R\'epublique, 92320 Ch\^atillon, France\\
Email: elie.cali@orange.com and taoufik.ennajjary@orange.com}
\IEEEauthorblockA{\IEEEauthorrefmark{1}Universit\'e Pierre et Marie Curie, 4 Place Jussieu, 75005 Paris, France
\\ Email: nila.novita.gafur@gmail.com}
\IEEEauthorblockA{\IEEEauthorrefmark{3}Mathematisches Institut, Ludwig-Maximilians-Universit\"at M\"unchen, Theresienstra\ss e 39, 80333 Munich, Germany\\Email: hirsch@math.lmu.de}
\IEEEauthorblockA{\IEEEauthorrefmark{4}Weierstrass Institute for Applied Analysis and Stochastics, Mohrenstra\ss e 39, 10117 Berlin, Germany\\
Email: benedikt.jahnel@wias-berlin.de and robert.patterson@wias-berlin.de}
}

\maketitle

\begin{abstract}
We study fundamental characteristics for the connectivity of multi-hop D2D networks. Devices are randomly distributed on street systems and are able to communicate with each other whenever their separation is smaller than some connectivity threshold. We model the street systems as Poisson-Voronoi or Poisson-Delaunay tessellations with varying street lengths.
We interpret the existence of adequate D2D connectivity as percolation of the underlying random graph.
We derive and compare approximations for the critical device-intensity for percolation, the percolation probability and the graph distance.
Our results show that for urban areas, the Poisson Boolean Model gives a very good approximation, while for rural areas, the percolation probability stays far from 1 even far above the percolation threshold. 
\end{abstract}

\IEEEpeerreviewmaketitle

\section{Introduction}
\label{sec:Intro}
In recent years, network models featuring device-to-device (D2D) communications have evolved into a vigorous research field, covering topics as broad as security, interference, energy consumption, radio communications and resource allocation, \cite{d2da, d2db}. However, the problem of the connectivity of wireless multi-hop networks has received little attention beyond highly simplified mathematical models. In particular, linking connectivity problems to the geometry of the underlying street system of the territory under consideration is new and an important step towards a more realistic analysis of D2D networks. As has been observed in~\cite{Glau}, street systems can heavily influence connectivity properties of the D2D network since, for example, thresholds for percolation are sensitive to the spatial positioning of the nodes.

For applications in telecommunications, Cox point processes are commonly employed to model various kinds of networks~\cite[Chapter 5]{Spor}, because Cox point processes model devices distributed randomly on environments which are themselves given by a random process. A variety of tessellation types have been proposed for this purpose including Poisson-Voronoi tessellations (PVT),  Poisson-Delaunay tessellations (PDT), line tessellations, and even more involved models like the nested tessellations in~\cite{Maier}. As a rough guideline, PDT fits better to rural areas since it represents roads between cities, whereas PVT is a useful model for different urban environments. Authors in~\cite{Glo1} present a detailed discussion of these tessellations and statistical comparisons to real city layouts.
Indeed, these random tessellations capture the statistical properties of real city layouts rather well using a small parameter space. 

Once the street system is given, devices are randomly positioned on the streets, either to model nodes of a fixed network, like distribution units, or mobile users positioned along the streets. The simplest way to do that is to use a linear Poisson point process confined to the random streets, i.e., a Cox point processes.

In case of a fixed network, the distribution of distances between nodes following the street system can be expressed via Palm theory~\cite{Glo2}. When deploying a hierarchical network, this entails the distribution of attenuation for signals travelling between customers and their access points. In particular, estimates for the deployment costs of the network on a given territory can then be derived~\cite{NTS}.
In the case of mobile networks, building the Gilbert graph (i.e., drawing an edge between any two devices with separation less than a given connection radius) one obtains a simple model for devices using D2D communications. 

\begin{figure}
\begin{center}
\input{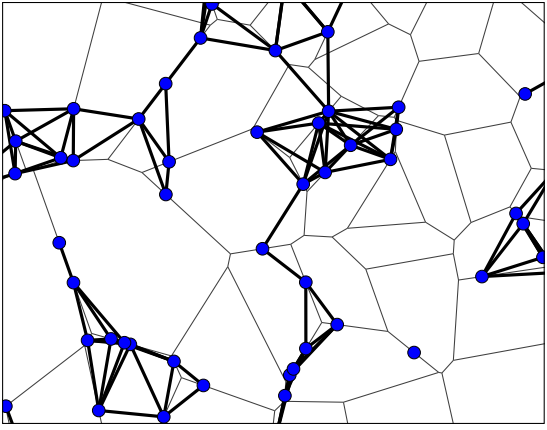}
\caption{The network model as a Gilbert graph: streets (thin lines) are represented by a PVT; devices (dots) are positioned uniformly at random on the streets; edges (thick lines) are drawn whenever two devices are separated by less than a connectivity threshold $r>0$.}
\label{fig:model}
\end{center}
\end{figure}

Such a network is only useful if a high proportion of users are connected to each other, even if they are on opposite sides of the city.  In other words, the Gilbert graph should feature a large connected component reaching all, or at least most, of the city.  Regarding a city as so large as an unbounded plane we interpret the desired connectivity as percolation  of the random graph~\cite{meesterRoy}.
Then, studying the percolation of this random graph, as depicted in Figure~\ref{fig:model}, one can obtain results on the connectivity of the wireless D2D network.

Analytic results are only available in the limiting regime of highly dense or very sparse street systems. In this paper, we give sufficiently precise numerical answers to the following three main questions:
\begin{enumerate}
\item What is the minimum density of devices in order to ensure long-distance communications?
\item What is the probability for a randomly selected device to be in the large connected component of the network?
\item How many hops per unit of distance are necessary to connect two devices in the same connected component?
\end{enumerate}

The network model used for the simulations is described in Section~\ref{sec:Model}. The simulation methods are described in Section~\ref{sec:Method}. Results and discussion follow in Section~\ref{sec:Results}.

\section{Network Model}
\label{sec:Model}

Following~\cite{Courtat, Stoyan}, we model a street system $T$ as a PVT or PDT based on a planar Poisson point process. Note that, with reference to the conditions presented in~\cite{Cox}, these street systems are stabilising and asymptotically essentially connected. In particular, we do not consider Poisson line tessellations, which are not stabilising. We denote by $\gamma = \E[\nu_1(T \cap [-1/2, 1/2]^2)]$ the length intensity of $T$ where $\nu_1$ measures the one-dimensional total length of $T$ in the unit square $[-1/2, 1/2]^2$. For a street system in a given city or part of a city, it is possible to obtain the best fitting tessellation process and the corresponding value of $\gamma$ easily using the statistical tools presented in~\cite{NTS}. 

For any realisation $T$ of the street system, devices are distributed according to a one-dimensional Poisson point process $X$ with intensity measure $\lambda \nu_1(T\cap d u)$ where $\lambda>0$ is a scalar parameter. In other words, the number of devices in an area $A$ is a Poisson random variable with parameter $\lambda \nu_1(T\cap A)$ and each device is placed independently and uniformly in $T\cap A$. In short, $X$ is a Cox point process  with random intensity measure $\lambda \nu_1(T\cap du)$.

As in~\cite{Glau}, we assume the communication radius between devices to be constant, so that we can construct the communication graph as a Gilbert graph with radius $r>0$. The answer to the first question presented in the introduction then amounts to finding the critical percolation threshold $\lambda_c$ of this Gilbert graph.

Answering the second question amounts to tracing the percolation function $\lambda\mapsto\theta(\lambda,r,\gamma)$ 
where 
$$\theta(\lambda,r,\gamma) = \P(o \rightsquigarrow\infty)$$
is the Palm probability, that a randomly selected user is connected to infinity in the Gilbert graph.

Finally, in view of sub-additivity of the number of hops and the corresponding results on the Poisson Boolean model (PBM), see~\cite{stretch}, it is reasonable to assume the
existence of a deterministic stretch factor $\mu(\lambda, r, \gamma) > 1/r$ given by 
$$\lim_{\substack{X_i, X_j \rightsquigarrow \infty \\ |X_i - X_j|\to\infty}} \frac{S(X_i, X_j)}{|X_i-X_j|} = \mu(\lambda, r, \gamma).$$
Here, $S(x,y)$ denotes the smallest number of hops from $x$ to $y$ and we assume $\lambda>\lambda_c$.
%
%
Therefore, the answer to the third question will consist in tracing the curves $\lambda\mapsto\mu(\lambda,r,\gamma)$ for given $r$ and $\gamma$.

The model presented here exhibits a scaling invariance: Changing $\gamma$ to $a \gamma$ is equivalent to zooming to a larger or smaller window for the original $\gamma$ value, thus changing $\lambda$ to $a\lambda$ and $r$ to $r/a$. Consequently, we considered only two dimensionless parameters, namely $\lambda/\gamma$ and $r \gamma$, for the simulations, and will switch back to three parameters when necessary in order to present the results in physical terms.

While the inclusion of a statistically realistic model for the city layout is a major step forward, three important simplifying assumptions are still required:
\begin{enumerate}
    \item The transmission power of devices is a global constant.
    \item Signal shadowing and fading effects due to the urban landscape are neglected.
    \item Interference between devices is not taken into account.
\end{enumerate}

\section{Methodology}
\label{sec:Method}
In order to answer the three key questions raised in Section~\ref{sec:Intro}, we rely on Monte Carlo estimates of three key quantities: the critical percolation threshold, the percolation probability and the stretch factor. First, the critical percolation threshold is needed to decide whether long-range multi-hop communication is possible at all. Second, the percolation probability describes the proportion of devices in an unbounded connected component : this gives the probability for a given device to be able to communicate on a long-range scale. Third, the stretch factor determines how many hops are needed to cover a given large distance.

\subsection{Critical percolation thresholds}

\begin{figure}
\begin{center}
\includegraphics[scale=0.55]{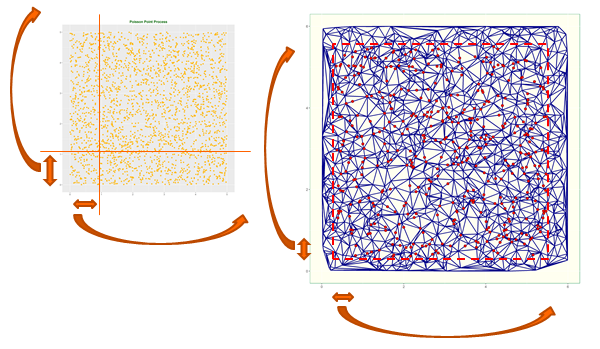}
\caption{Tracing on a torus : after plotting a planar Poisson process in a $5{\rm km}\times 5{\rm km}$ window, the points on the left are copied to the right and the points at the bottom are copied to the top. The corresponding PDT is traced and the dotted line in red thus delimits a $5{\rm km} \times 5{\rm km}$ window representing the torus. The devices (red dots) could then be placed uniformly at random on the edges of the graph inside the window, and the devices in a band on the left or at the bottom are copied to the right and top.}
\label{fig:torus}
\end{center}
\end{figure}

The first problem we encounter with simulations is to estimate infinite connected components in finite windows. To tackle this problem, we realize the tessellations on a torus. Thereby the left side and the right side of the window as well as the up and bottom sides could adjust perfectly; see Figure~\ref{fig:torus}. In order to do that, we simply generate a Poisson point process in the finite window, and then copy the points from a large band on the left of the window to the right, and correspondingly from the bottom to the top of the window. 
In order to avoid boundary effects, we then generate devices uniformly at random on the streets in a sub-window in the center of the extended window, and copy the devices from the left and bottom sides of the window to the right and up sides. Infinite connected components then correspond to components that cross the window, either from left to right or from top to bottom, joining a user to its copy.

\begin{figure}
\begin{center}
\includegraphics[scale=0.4]{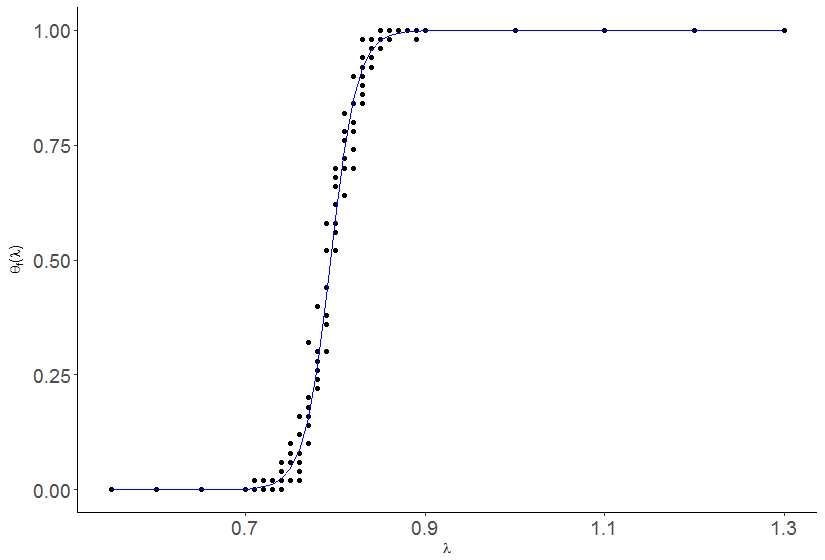}
    \caption{Proportion of simulations where we find an infinite component as a function of $\lambda$. Each point represents an estimate for a given $\lambda$. The blue line represents the logistic regression curve. }
\label{fig:confidence}
\end{center}
\end{figure}

\begin{figure}
\begin{center}
\includegraphics[scale=0.12]{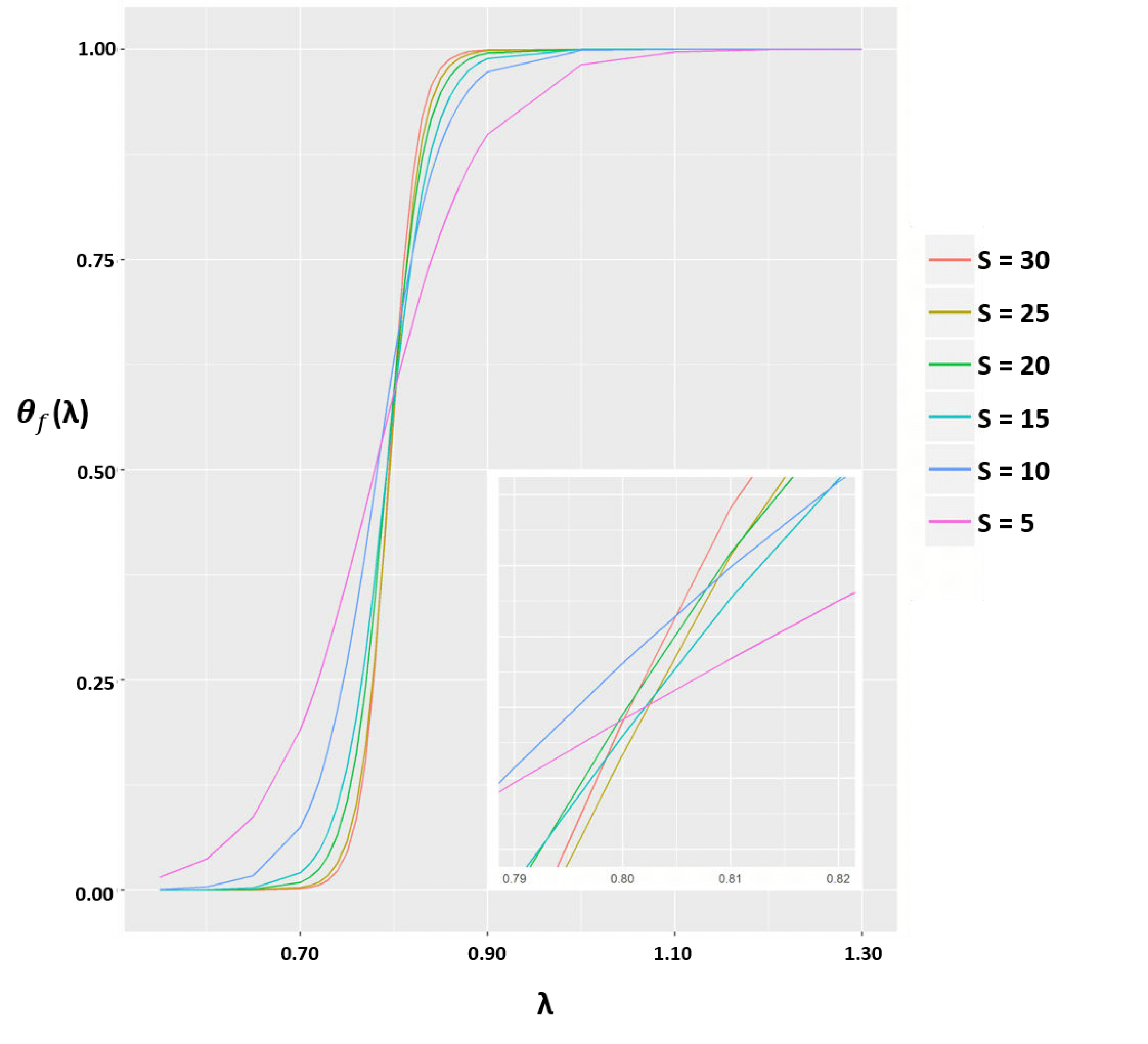}
    \caption{Tracing curves $\lambda\mapsto p(\lambda)$ for window sizes varying between $5 {\rm km}\times 5{\rm km}$ and $30{\rm km}\times 30{\rm km}$: the curves intersect near $p = 0.6$.}
\label{fig:intersect}
\end{center}
\end{figure}

Figure \ref{fig:confidence} shows the estimate of the proportion of simulations $p$ where we find an infinite component as a function of $\lambda$.  The logistic model $\log(p/(1-p)) = a\lambda + b$ seems to perfectly fit the curve. Finding $a$ and $b$ by linear regression, it is then possible to obtain the value of $\lambda$ for a given $p$.
Figure~\ref{fig:intersect} presents the logistic curves for varying window sizes. First, we observe that as the simulation windows grow, the curves tend to the theoretical function of the percolation probability (in the limit when the window's size goes to infinity, the $p(\lambda)$ function is $0$ for $\lambda<\lambda_c$ and $1$ for $\lambda>\lambda_c$). Second, we observe that all the curves cross in a limited region around $p = 0.6$, thus, we approximate  $\lambda_c$ via $$\lambda_c \approx (\log(0.6/0.4) - b)/a.$$


\subsection{Percolation probabilities}\label{PerProb}

In order to calculate the percolation probabilities we use the following algorithm. First, for a realisation $t$ of the random tessellation, we start by selecting the origin $o$ uniformly from the street system. Then, we add points sequentially on the street system, again following the uniform distribution. For each new point, we use the union-find algorithm, see~\cite{UF}, in order to update the connected components of the Gilbert graph. We stop adding points as soon as the connected component containing the origin crosses the window from a point to its copy. For each tessellation $t_k, k=1,\dotsc, n$ we run the simulation of point placements $M$ times and for the $i$th such simulation we denote by $N_{k,i}$ the number of points needed to exhibit the window crossing.

Let $\{o\rightsquigarrow\partial W\}$ denote the event that zero is contained 
in a connected component which crosses the window $W$. Further, let $J$ denote the number of points of the process $X$ of devices in the simulation window and let 
$$\theta_t = \sum_{j>0} \P(o \rightsquigarrow\partial W|J=j, T=t) \P(J=j | T=t)$$
so that the expected value of $\theta_T$ (with respect to the tessellation process $T$) is a finite-volume estimator of $\theta$.

For given a realisation of the street system $T=t$ the number of devices is Poisson distributed with mean $\lambda \nu_1(t)$ and hence
$$\P(J=j | T=t) = \exp(-\lambda \nu_1(t)) \frac{\lambda^j\nu_1(t)^j}{j!}.$$
Moreover, since each tessellation realisation is assigned equal weight and since the property of percolation is stable with respect to the addition of further devices, a natural estimator for $\theta$ is

\begin{equation}\label{estimator}
\hat{\theta}(\lambda) = \frac{1}{n} \sum_{k \le n}\frac{1}{M} \sum_{i \le M} \sum_{j \ge N_{k,i}} e^{-\lambda \nu_1(t_k)}\frac{\lambda^j\nu_1(t_k)^j}{j!}.
\end{equation}

The sum $\sum_{j \ge N_{k,i}} e^{-\lambda \nu_1(t_k)}\frac{\lambda^j\nu_1(t_k)^j}{j!}$ equals $\P\left(J \geq N_{k,i} \mid T = t_k\right) $ and is therefore simply one minus the cumulative distribution function for a Poisson random variable with known mean and thus readily available from computational libraries.
Note that each simulation will give a value for $N_{k,i}$, and then values of $\hat \theta$ for different $\lambda$ can be computed from \eqref{estimator} without extra simulations.

\subsection{Stretch factor}

Since we must have an infinite connected component, we consider only values $\lambda > \lambda_c$. In order to compute the stretch factor for each $\lambda$ value, we first determine the chemical distance, i.e., the minimal number of hops between two devices for all pairs in a given simulation via the Floyd-Warshall algorithm~\cite{FW}. Then, we compute the Euclidean distance between each pair of points. Since $\mu$ is a limit for large distance only pairs with distance more than $4{\rm km}$ in a window of side length $5{\rm km}$ are taken into account for the estimation.

\section{Results}
\label{sec:Results}
We now present the results of the Monte Carlo experiments described in the previous sections.
Simulations were performed in the statistical computing environment R \cite{RManual}.  Unless stated differently, the street system length intensity is fixed to $\gamma = 20 {\rm km}^{-1}$ which corresponds to dense urban areas.

\subsection{Estimation of the percolation threshold}

The observation window for the simulations is fixed to $30{\rm km}\times 30{\rm km}$ unless stated otherwise. For certain parameter combinations the window size had to be reduced due to computational constraints.

\begin{table}
\begin{center}

\caption{Percolation threshold $\lambda_c/\gamma$ as function of $r \gamma$.}
\label{tab:lambda_gamma20}
\begin{tabular}{|c m{1.5cm} m{1.5cm} m{1.5cm}|}
\hline
        & \multicolumn{3}{c|}{$\lambda_c/\gamma$}  \\
$r\gamma$ & PVT & PDT & PBM  \\
\hline
0.3 & 11.9 & 12.4 & 15.95 \\
0.5 & 5.58 & 5.83 & 5.74\\
1.5 & 0.75 & 0.79 & 0.638\\
2.5 & 0.24 & 0.26 & 0.229\\
3.5 & 0.12 & 0.13 & 0.117\\
4.5 & 0.071 & 0.075 & 0.0709\\
5.5 & 0.048 & 0.049 & 0.0474\\
6.5 & 0.034 & 0.035 & 0.034\\
7.5 & 0.025 & 0.026 & 0.0255\\
8.5 & 0.0199 & 0.0202 & 0.01995\\
9.5 & 0.0159 & 0.0159 & 0.0159\\
\hline

\end{tabular}
\end{center}
\end{table}

%

Table~\ref{tab:lambda_gamma20} shows the various critical values found for $\lambda/\gamma$, allowing to compute $\lambda_c$ for various values of $\gamma$ for PDT and PVT models, to be compared to the standard Poisson Boolean Model (PBM). The simulations for $r \gamma = 1.5$ were performed on a $15{\rm km}\times 15{\rm km}$ window, those for $r \gamma = 0.5$ and those for $r \gamma = 0.3$ on a $10{\rm km}\times 10{\rm km}$ window. Note that a comparison with the critical intensity for the PBM leads to the approximation
$$\lambda_c\gamma^{-1}\approx 4.51\pi^{-1}(r\gamma)^{-2},$$
see~\cite[Table 2]{Ballister}. 

For a standard city centre, $\gamma\approx 20{\rm km}^{-1}$, so that if $r = 0.475{\rm km}$, we have that $\lambda_c = 0.32{\rm\ devices/km}$, which means that the density of devices per square kilometre needs only to be greater then $\lambda_c \gamma = 6.4{\rm\ devices/km^2}$, while if $r = 0.075{\rm km}$, it should be greater than $300{\rm\ devices/km^2}$.
For a rural area, $\gamma \approx 1 {\rm km}^{-1}$, if $r = 0.3{\rm km}$, the number of devices per kilometre of streets must be greater than $12$ to have percolation.

We can see here that the results are very near to the PBM for high values of $r \gamma$, which means that in the limit of dense streets with a reasonable interaction radius, the street system does not play a role and the Cox point process is near to a planar Poisson point process. 
On the other hand, when the street system becomes sparse, for low values of $r \gamma$, the PBM is not a good approximation. An inhomogeneous Bernoulli bond percolation model can be used for approximations, for details see~\cite{Cox}. In the case of PVT, the critical Bernoulli parameter can be approximated by $b_c = 0.5$, resulting in the approximation
$$\lambda_c/\gamma \hbox{  }\exp(-r \lambda_c) = -\log(b_c).$$
This gives $\lambda_c/\gamma \approx 12{\rm\ devices}$ which is very close to the value found for $r \gamma = 0.3$ in the simulations.

The simulations also show that the transition from the approximation by the PBM to the approximation by the inhomogeneous Bernoulli bond percolation is very steep, which means that almost all values of $r\gamma$ can be covered by one of the two approximations presented. This also gives us the domain of validity for each approximation.

\subsection{Estimation of the percolation probability}

\begin{figure}
\begin{center}
\includegraphics[scale=0.35]{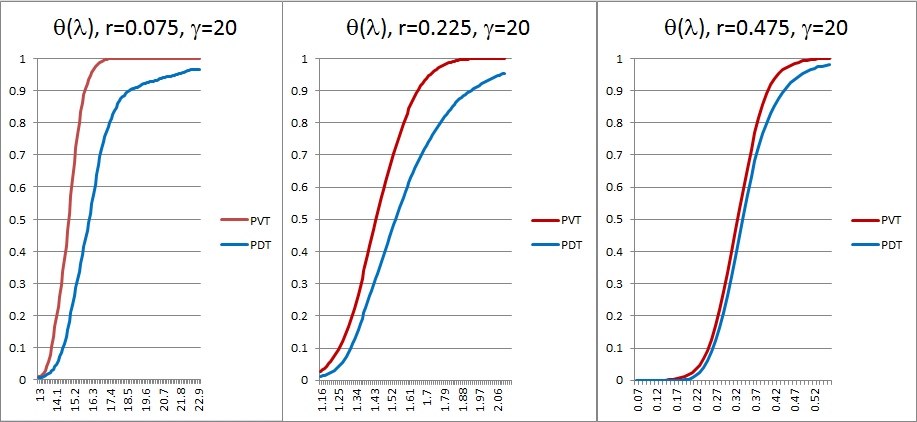}
    \caption{Percolation probability as a function of intensity for varying communication radii.}
\label{fig:thetaPVTPDT}
\end{center}
\end{figure}

The simulation method for the percolation probability $\theta$ follows the algorithm described in Section~\ref{PerProb}. Again, the tessellation is traced and the devices placed on a torus. For each value of $r\gamma$, the estimates are based on $N = 10$ realisations of the random tessellation and $M = 30$ realisations of randomly placed devices on each of these tessellations. The results are presented in Figure~\ref{fig:thetaPVTPDT}. The $\theta$ curve gives the probability for a randomly selected user to belong to the infinite connected component. The curves have been calculated for various values of $r\gamma$. 
Thus, for big interaction radii, both PDT and PVT model are very close to the standard PBM, while for smaller radii, the percolation probability is always smaller for PDT than for PVT.

The curves show the significant difference appearing between PDT and PVT models when $r\gamma$ becomes small. In the PDT model, the percolation probability $\theta$ increases much more slowly. In particular, even far above the percolation threshold, some devices are not connected to the network.

\subsection{Estimation of the stretch factor}

\begin{figure}
\begin{center}
\includegraphics[scale=0.16]{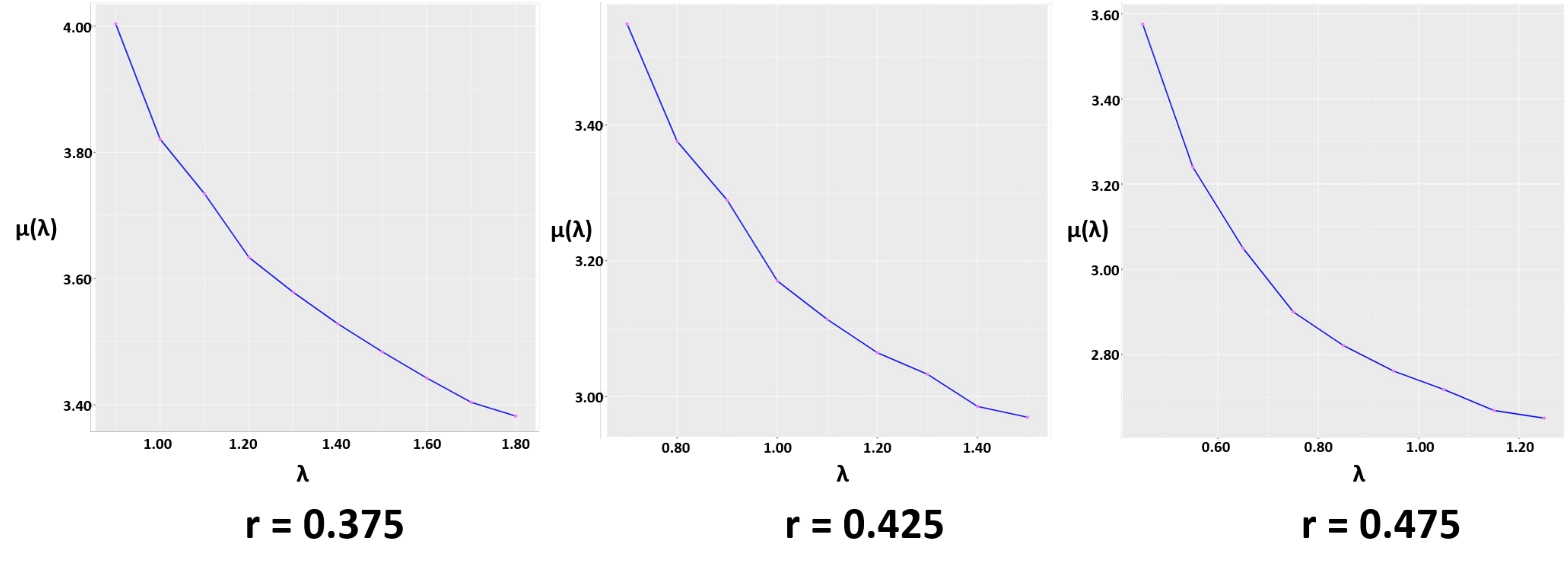}
    \caption{Stretch factor for PVT as a function of intensity for varying communication radii.}
\label{fig:muPVT}
\end{center}
\end{figure}

\begin{figure}
\begin{center}
\includegraphics[scale=0.16]{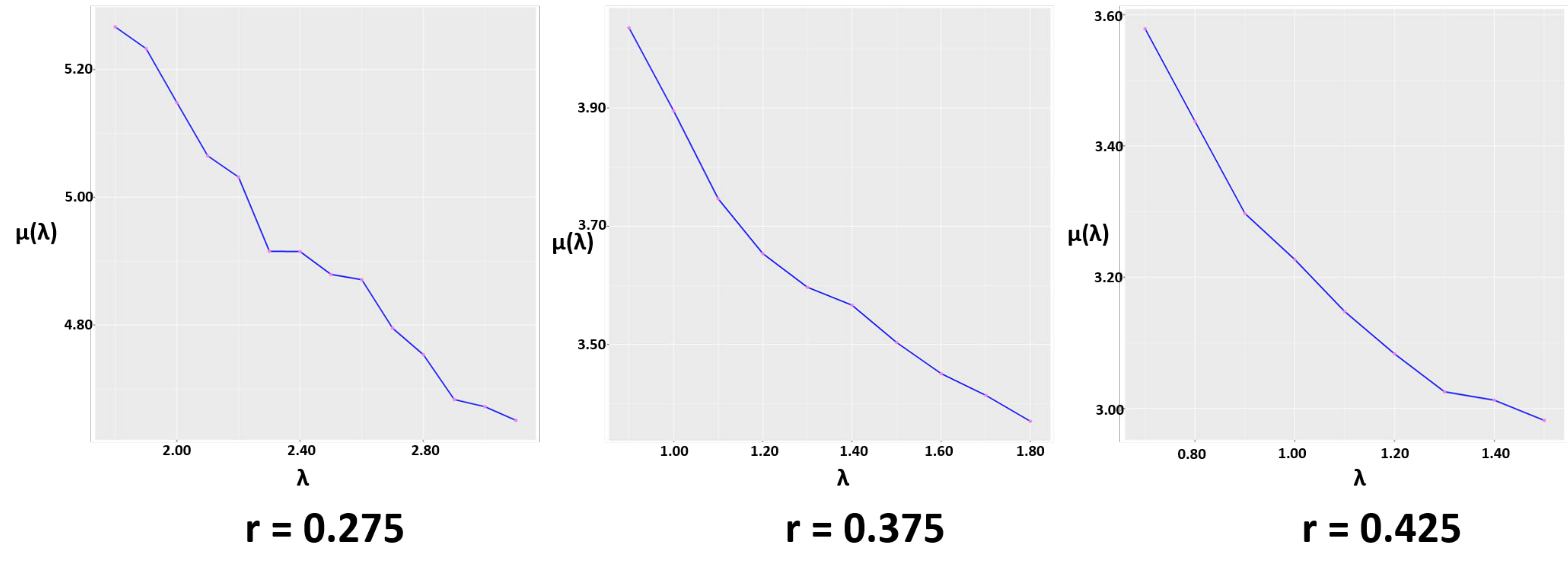}
    \caption{Stretch factor for PDT as a function of intensity for varying communication radii.}
\label{fig:muPDT}
\end{center}
\end{figure}

Here, the window is $5{\rm km}\times 5{\rm km}$ and again the tessellations and devices are traced on a torus. In order to estimate the stretch factor $\mu$, the values of $\lambda$ are taken above the critical value, and only devices in the infinite connected component are taken into account. 
For each value of $\lambda$, $100$ simulations are performed. Figures~\ref{fig:muPVT} and~\ref{fig:muPDT} illustrate the results with $\gamma = 20{\rm km}^{-1}$ and varying $r$. 
      For instance in PVT with $r = 0.375{\rm km}$ and $\lambda = 1.5 {\rm\ devices/km}$, if two devices in the infinite connected component are at distance $2{\rm km}$, then approximately $7$ hops are required. Additionally, note that the stretch factor is bigger for PDT than for PVT.

\section{Conclusion}
The results obtained in this paper are a first step towards analysing connectivity in multi-hop wireless networks. For rural areas, where shadowing and interferences have little impact, the model already covers a large proportion of pertinent features. It reveals that the inhomogeneous Bernoulli bond percolation model gives a much better approximation for these type of areas than the PBM. Even more, it shows that in rural areas, for PDT models, the percolation probability will stay far from $1$. Although realistic models for urban areas will need to take into account the propagation effects caused by the urban landscape, the proposed model already gives strong hints on methods for addressing the basic problem of connectivity.

Dealing with shadowing in urban areas now seems to be a reasonable next step. One idea that we are currently exploring is to introduce a reduced connection radius for points that do not belong to the same street segment.
Moreover, other street system models, like Manhattan grids, or embedded tessellations could also be studied in order to represent a larger class of real world street systems.

\section*{Acknowledgments}
This research was supported by the Leibniz program \emph{Probabilistic Methods for Mobile Ad-Hoc Networks}, by the LMU Munich's Institutional Strategy LMUexcellent within the framework of the German Excellence Initiative and by  Orange  S.A.~grant  CRE G09292.

\bibliographystyle{IEEEtran}
\bibliography{pres_shanghai}

\end{document}